\documentclass[conference]{IEEEtran}
\IEEEoverridecommandlockouts
\usepackage{cite}
\usepackage{subfigure} 
\usepackage{amsmath,amssymb,amsfonts}
\usepackage{algorithmic}
\usepackage{graphicx}
\usepackage{textcomp}
\usepackage{xcolor}
\usepackage{multirow}

\usepackage{fancyhdr}
\fancypagestyle{firststyle}
{
   \fancyhf{}
   \fancyhead[l]{The published version is available at: \url{https://doi.org/10.1109/ICDCS47774.2020.00165}}
}

\def\BibTeX{{\rm B\kern-.05em{\sc i\kern-.025em b}\kern-.08em
    T\kern-.1667em\lower.7ex\hbox{E}\kern-.125emX}}
\begin{document}

\title{Performance Characterization and Bottleneck Analysis of Hyperledger Fabric}

\author{\IEEEauthorblockN{Canhui Wang, Xiaowen Chu}
\IEEEauthorblockA{Department of Computer Science, Hong Kong Baptist University \\
Email: \{chwang, chxw\}@comp.hkbu.edu.hk}
}
\maketitle

\begin{abstract}

Hyperledger Fabric is a popular open-source project for deploying permissioned blockchains. Many performance characteristics of the latest Hyperledger Fabric (e.g., performance characteristics of each phase, the impacts of ordering services, bottleneck and scalability) are still not well understood due to the performance complexity of distributed systems. We conducted a thorough performance evaluation on the first long term support release of Hyperledger Fabric. We studied the performance characteristics of each phase, including execute, order, and the validate phase, according to Hyperledger Fabric’s new execute-order-validate architecture. We also studied the ordering services, including Solo, Kafka, and Raft. Our experimental results showed some findings as follows. 1) The execution phase exhibited a good scalability under the OR endorsement policy but not with the AND endorsement policy. 2) We were not able to find a significant performance difference between the three ordering services. 3) The validate phase was likely to be the system bottleneck due to the low validation speed of chaincode. Overall, our work helps to understand and improve Hyperledger Fabric.

\end{abstract}


\thispagestyle{firststyle}
\section{Introduction}
Blockchain is an immutable distributed ledger system, originally proposed by Satoshi Nakamoto to solve the double-spending problem in Bitcoin\cite{nakamoto2019bitcoin}. People can view blockchain as a chain of blocks with important new features (e.g., open, immutability, and integrity) which allows users to issue and clear transactions in a decentralized fashion without relying on a trusted third party. Blockchain creates a new way of exchanging digital goods and values. One of the most popular applications that benefit from the Blockchain technology is cryptocurrency, such as Bitcoin and Ethereum. Due to the ability to provide record keeping and the irrevocability of data sharing, blockchain technology has been showing great potential in various areas, including retails, healthcare, and financial applications.

Hyperledger Fabric (HLF) is a popular open-source project for deploying permissioned blockchains. Permissioned blockchains are the blockchains where permission is required to become a member of the blockchain network. There are some restrictions on who can read and write the blockchain data. A permissioned blockchain grants access to those identified participants to define who has the read permission or write permission. For example, people can establish a permissioned blockchain between three or more organizations, where some participants can read and write transactions, and others can only read the transaction data. One of the most popular permissioned blockchain platforms \cite{androulaki2018hyperledger,thakkar2018performance} is Hyperledger Fabric.

Performance characterization and bottleneck analysis in Hyperledger Fabric are important for three main reasons. Firstly, a performance study on a newer version of Hyperledger Fabric is vital because it usually provides new features and optimizations. However, many recent works \cite{androulaki2018hyperledger, thakkar2018performance} are based on Hyperledger Fabric version 1.1.x or older. Secondly, an in-depth understanding of each phase of Hyperledger Fabric is essential because it helps to understand system bottlenecks better. However, some benchmark tools such as Hyperledger Caliper did not provide an in-depth performance analysis of each phase, and there are few related research work. Thirdly, a performance comparison of the three ordering services is important because it helps to better understand the performance characteristics of ordering services (including Solo, Kafka, and Raft) in the context of Hyperledger Fabric. Although a few related works \cite{baliga2018performance, klaokliang2018novel, baliga2018performance} are available, there lacks a comprehensive comparison. To solve these problems, we conducted experiments on Hyperledger Fabric v1.4.3 LTS and aimed to provide an in-depth performance analysis on each phase. Our contributions can be summarized as follows.

\begin{itemize}
    \item We analyzed the performance characteristics of each individual phase, following Hyperledger Fabric’s new execute-order-validate architecture.    
    
    \item We conducted a performance comparison on different ordering services, and found no significant difference among the Solo, Kafka, and Raft settings, in the context of Hyperledger Fabric. 
    
    \item We studied the scalability of the endorsing peers and the ordering service nodes. For example, we found that the execute phase showed a better performance scalability under the OR endorsement policy, but it becomes exacerbated under the AND endorsement policy.
    
    \item The system bottleneck of our experiments was in the validate phase because the speed of validation of blocks (and transactions) is slow.
    
\end{itemize}

The rest of the paper is organized as follows. Section \uppercase\expandafter{\romannumeral2} introduces the core components of Hyperledger Fabric. Section \uppercase\expandafter{\romannumeral3} presents the preliminaries of the ordering services. Section \uppercase\expandafter{\romannumeral4} presents our experimental setting and analyzes the experimental results. Section \uppercase\expandafter{\romannumeral5} introduces related works in the performance evaluation of Hyperledger Fabric. Finally, section \uppercase\expandafter{\romannumeral6} concludes the paper.

\section{Hyperledger Fabric}

\textit{Hyperledger Fabric} is a popular permission blockchain platform that provides various pluggable modular components such as membership service provider \cite{park2018totp}, chaincode \cite{androulaki2019endorsement}, and ordering service \cite{sousa2018byzantine}. Unlike permissionless blockchains, each participant of HLF must be identified by the fabric certificate authority. Each transaction is processed on a predefined channel where participants are authenticated and authorized to do specific operations such as read, write, and endorse on transaction data according to the channel configuration settings. These transaction data and channel configuration settings are recorded in the HLF ledger.

\textit{Fabric Certification Authority (Fabric CA)} is an identity management system \cite{mazumdar2019design}. It issues enrolment certificates to the major participants of the Fabric network, including ordering service nodes, peers nodes, and client nodes.

\thispagestyle{empty}
\textit{Ordering Service Nodes (OSNs)} are the nodes that collectively form the ordering service of Hyperledger Fabric \cite{bessani2017byzantine}. The ordering service receives transactions from all channels in the network, orders them chronologically on a per-channel basis, and packages them into blocks. The blocks will be delivered to peers on the channel for final validation and commitment. This design choice renders consensus in Fabric as modular as possible and simplifies the replacement of ordering service nodes.

\textit{Peer Nodes} are the nodes \cite{androulaki2018hyperledger} that endorse transaction proposals, validate and commit transactions. A subset of them called endorsing peers endorses the transaction proposals. The endorsing peers verify the following. 1) The transaction proposal is well-formed. 2) The transaction has not been submitted in the past. 3) The signature is valid. 4) The submitter of the transaction proposal is authorized to transact on the channel. If all checks go through, then the chaincode is executed, and proposal responses are generated and signed by the endorsing peers and sent back to the client for the next ordering phase. The second operation is to validate and commit transactions. All peers of the channel have to validate and commit all transactions of a block. The peers validate that the endorsement policy is fulfilled, and a read-write conflict check ensures that there have been no changes to ledger state for read set. If all validations go through, transactions in the block are tagged as valid; otherwise, it will be tagged as invalid. Both valid and invalid transactions are recorded into the blockchain, while only valid transactions update the world state.

\textit{Client Nodes} are the nodes that can be defined to prepare transaction proposals, to collect proposal responses, and to submit it for the ordering phase \cite{thakkar2018performance}.

\textit{A Channel} is a private blockchain subnet \cite{androulaki2018channels} where specific network participants are allowed to conduct private and confidential transactions.

\textit{A Chaincode} is a code that implements a set of business or application logics \cite{beckert2018formal}. Typically, these logics are agreed on by the participants of the network who have installed the chaincode. There are two types of chaincodes: user chaincodes and system chaincodes. User chaincodes run in a secured Docker container isolated from the peer process. System chaincode runs in the same process with the peer. Examples of system chaincodes include validation system chaincode (VSCC) – to validate a transaction’s endorsements, endorsement system chaincode (ESCC) – to endorse a transaction, and multi-version concurrency control chaincode (MVCC) – to avoid a double-spending attack and the same transaction cannot be accepted twice.

\textit{An Endorsement Policy} is a set of transaction validation rules that define necessary and sufficient conditions for a valid transaction endorsement \cite{benhamouda2019supporting}. A validation rule usually contains two parts: target endorsing peers and the Boolean operators. The first part dictates the required endorsements from a subset of peers. The second part supports arbitrary Boolean logic include AND, OR, and OUTOF. For example, a typical endorsement policy could specify a validation rule requiring at least $k$ endorsements from target $n$ endorsing peers.

\section{The Ordering Services}

\textit{Solo} is a solo implementation of the ordering service. It runs on a single-node mode, for which the single point of failure problem may occur. People usually used the solo mode for development and testing purposes.

\textit{Kafka} is a distributed implementation of the ordering service that provides consensus agreements and achieves crash fault-tolerant. The Kafka ordering service contains two main components: Broker and ZooKeeper. A broker is a node that runs the Kafka application. A ZooKeeper is a cluster that provides services such as leader election, membership management, and access control for the Kafka ordering service. Two core system parameters in Kafka need to be emphasized: partition and replication factor. A partition is a workspace where transactions are continually committed into the log in a specific order. In the Hyperledger Fabric context, a partition is a channel that maintains a sequence of well-ordered transactions. Kafka only provides the ordering service of transactions on a single partition. The replication factor is the number of partitions that should contain a copy of the partition log. A replication factor of $n$ means a partition is replicated $n$ times. A larger replication factor leads to higher system availability. Kafka follows a “leader and follower” model where if a broker by some reason is shut down, the ZooKeeper cluster will be responsible for selecting another partition replication as the data server. The process of log replication among replicas is called the in-sync replica. When Kafka receives a transaction from the user, it is written by the leader and replicated to the followers. A transaction is committed when it has been successfully replicated to all followers. This process may result in in-sync replica latency, in particular when the number of brokers is large enough. By default, our experiments set the partition parameter to 1 and the replication-factor parameter to 3.

\textit{Raft} is a distributed implementation of the ordering service. Raft provides consensus agreements that achieve crash fault-tolerant. It also follows the “leader and follower” model. The leader node’s decisions will be replicated to all followers nodes. Before committing the decision in the leader node, the leader needs to wait until a majority of follower nodes have successfully written the decisions. After committing the decision in the leader node, the leader node will notify all follower nodes to commit the decision, and the cluster has now come to a consensus state. Raft is a lightweight consensus framework that is easier to set up and manage. However, it provides fewer functionalities and tools to a distributed ordering service, such as partitions and replication factor.

There are two core conditions for the ordering service to generate a new block: BatchSize and BatchTimeout. BatchSize is the number of transactions per block. Our default setting of BatchSize is 100, which implies at most 100 transactions per block. BatchTimeout is the amount of time to wait before creating a new block. Our default setting of BatchTimeout is 1, which implies that at most waiting 1 second to generate a new block. For solo mode, the ordering service node simply cuts blocks whenever the number of transactions is greater than or equal to BatchSize, or the amount of time for BatchTimeout passed. For Kafka and Raft modes, the ordering service nodes provide an atomic block-cutting service. Similarly, the ordering service node cuts blocks whenever the number of transactions is greater than or equal to BatchSize, or a BatchTimeout Signal is received from the current leading node of the cluster.

\thispagestyle{empty}

\begin{figure}[b]
	\centering
	\scalebox{0.53}{
	\includegraphics[]{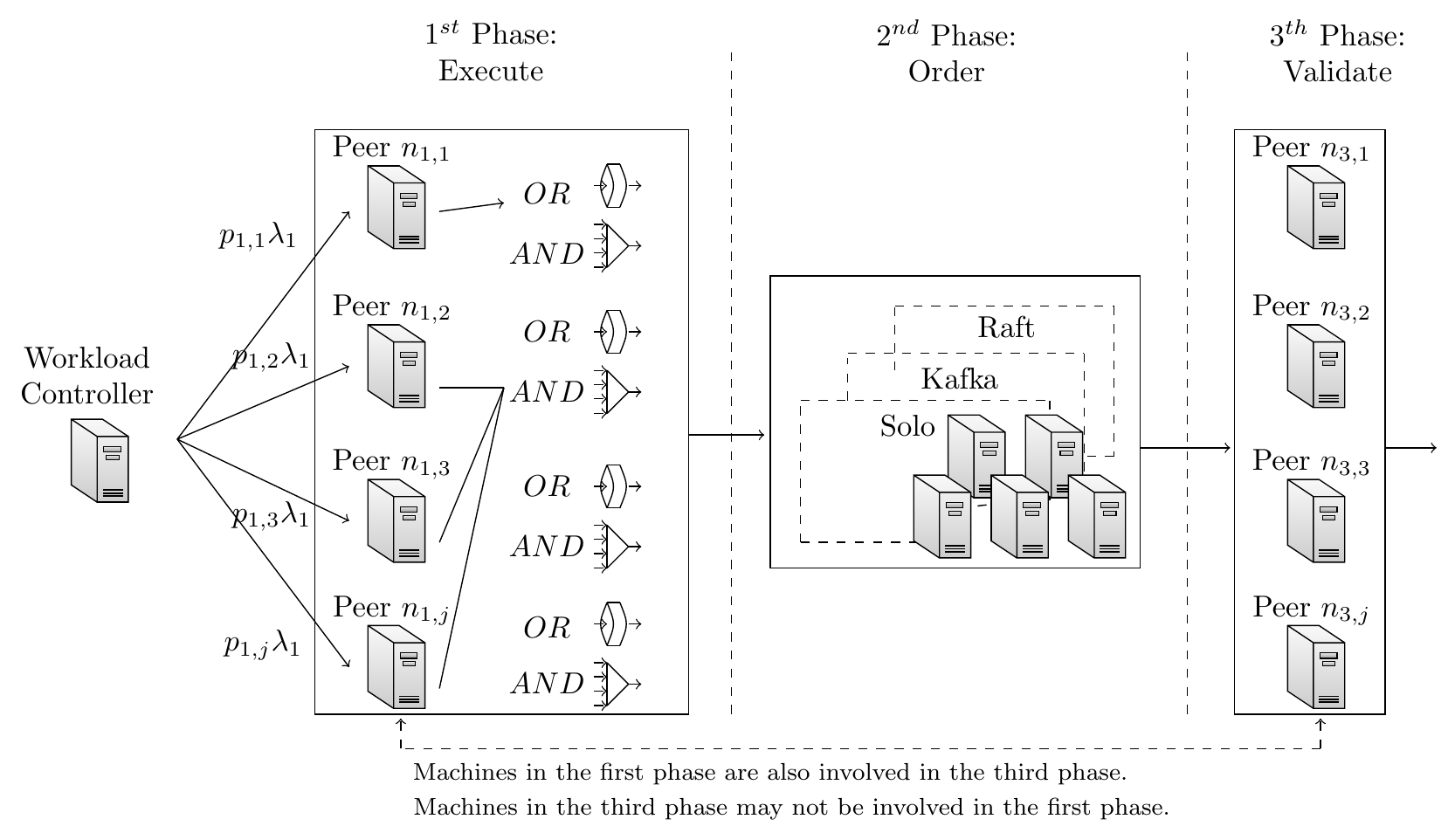}
	}
	\caption{System Architecture of Hyperledger Fabric}  
\end{figure}

\section{Performance Evaluation and Analysis}

\subsection{Experimental Setup}

Fig. 1 shows an overview of the system architecture of Hyperledger Fabric, where there are three phases: the execute phase, the order phase, and the validate phase from left to right, respectively. Specifically as shown in TABLE \uppercase\expandafter{\romannumeral1}, we used the Fabric v1.4.3 of the Hyperledger Fabric, which was the first long term run support of Hyperledger Fabric. We set up the Hyperledger Fabric network on our cluster of 20  machines. The 20 machines were inner-connect in a 1 Gbps network. The ordering service nodes were preferentially assigned to machines with Intel(R) Core(TM) i7-2600 CPU @ 3.40GHz. The endorsing peers were preferentially assigned to machines with Intel(R) Core(TM) i7-2600 CPU @ 3.40GHz. By default, our experiments will not endorse transactions from the ordering service nodes. For the software stack, we used Fabric SDK Node of version 1.0.0 and Nodejs of version 8.16.2. We enabled both the ordering service nodes and the peers node with transport layer security.

\begin{table}[t]
\centering
\caption{Experimental Configuration}
\scalebox{0.99}{
\begin{tabular}{|c|l|}
\hline
\begin{tabular}[c]{@{}c@{}}Environment\end{tabular} & Description                                      \\ \hline \hline
\multirow{2}{*}{CPU}                                                  & 8$\times$Intel(R) Core(TM) i7-2600 CPU @ 3.40GHz \\ \cline{2-2} 
                                                                      & 12$\times$Intel(R) Core(TM) i7 CPU 920 @ 2.67GHz \\ \hline
Memory                                                                & 4 GB DDR3                                        \\ \hline
Network                                                               & 1 Gbps Ethernet                                  \\ \hline
Hard Disk                                                             & SEAGATE ST3250310AS                              \\ \hline \hline
Hyperledger Fabric                                                    & Version 1.4.3                                    \\ \hline
Fabric-SDK-Node                                                       & Version 1.0.0                                    \\ \hline
Nodejs                                                                & Version 8.16.2                                   \\ \hline
\end{tabular}
}
\end{table}

We proposed several designs and principles in order to avoid the bottleneck occurs on the workload generator as follows. 1) We separated endorsing nodes and ordering service nodes, and ran these nodes on different machines. 2) We invoked multiple transactions per client. Because setting up a client requires some time-consuming MSP configurations. In the worst situation, if setting up one client each one transaction, it will require much computing overhead, meaning that each time when invoking a transaction, several time-consuming operations such as access configuration files locally will happen. It could result in a performance bottleneck. 3) We invoked transactions asynchronously. In our implementation, we used Nodejs asynchronous programming skills to invoke multiple transactions at the same time. We invoked new transactions without waiting for the responses of previous ones. 4) We used several clients as a workload generator simultaneously. Because using one machine is not able to overflow another machine with similar hardware settings. 5) We used a log system for double-checking that the load is generated or received at a specific rate.

\subsection{Performance Metrics}

We focused on transaction throughput and transaction latency. Following Hyperledger Fabric’s white paper, we defined the following performance metrics.

\textit{Definition 4.1 (Throughput).} The throughput is the rate at which transactions are committed to the ledger in transactions per second.

\textit{Definition 4.2 (Latency).} The latency is the committing timestamp of a transaction minus the submission timestamp of a transaction, where the committing timestamp means the timestamp when a transaction is committed to the ledger. We calculate the transaction latency for each transaction and get an average latency in total.

\textit{Definition 4.3 (Block Time).} Block time means how long the ordering service nodes take to create a new block on average. We can use logs at ordering service nodes to calculate the block time. Through block time, we can further calculate the throughput of the ordering service.

\subsection{Experimental Results}

\textbf{Overall Throughput.} Fig. 2 shows an overall transaction throughput under different ordering services and endorsement policies. We have two observations. First is that the maximum throughputs are around 300 tps under the endorsement policy of OR and transaction size of 1 byte. We found there is no significant difference in the maximum performance achieved by the three ordering services, though the last two provide useful features such as fault-tolerant, load balance, and replicas. Second is that the system performance in handling endorsement policy of AND is around 200 tps, which is significantly lower than OR. 

\begin{figure}[htbp]
	\centering
	\hspace{-7pt}\includegraphics[width=0.50\textwidth]{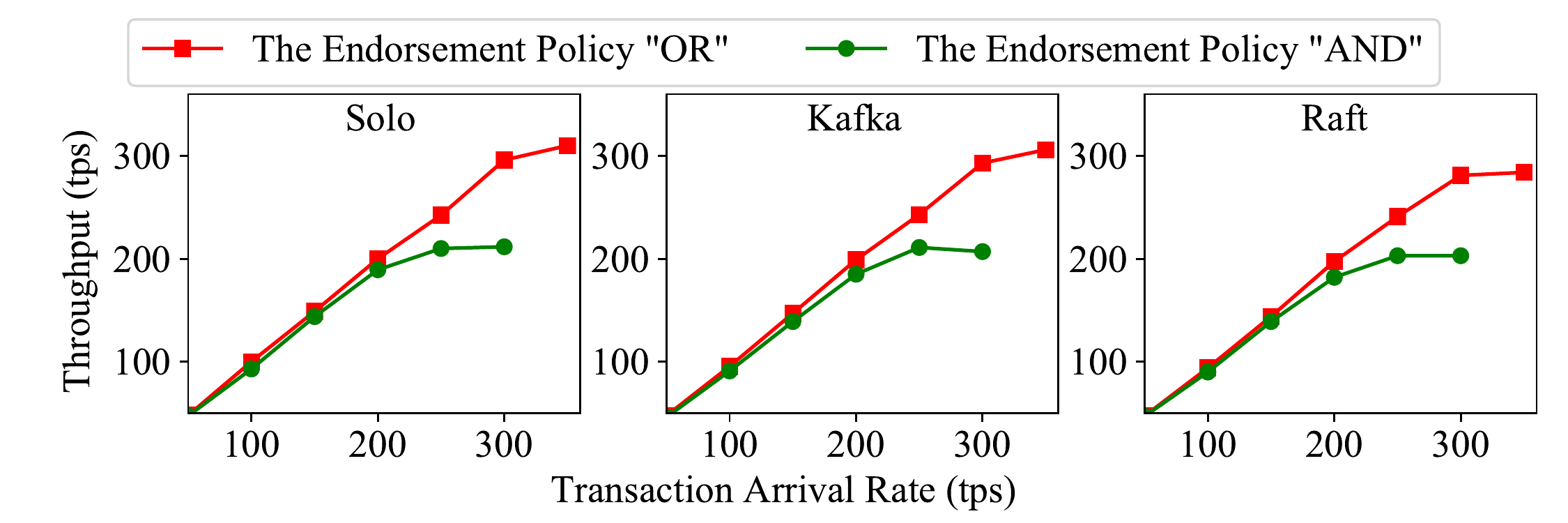}
	\caption{Overall Transaction Throughput}
\end{figure}

In summary, there is no difference in the maximum throughput achieved by the three ordering services. Also, the maximum throughput under the endorsement policy AND is significantly lower than OR.

 \thispagestyle{empty}
\textbf{Overall Latency.} Fig. 3 shows an overall transaction latency under different ordering services and endorsement policies. One observation is that when the system reaches its peak performance, the transaction latency will increase rapidly. In particular, the transaction latency with the endorsement policy AND increases more significantly than OR. Because the maximum throughput of AND is significantly lower than OR, and the performance bottleneck comes earlier than OR. Note that in our experiment, we used Nodejs and set the maximum transaction latency for the ordering service 3 seconds, meaning that if a client failed to receive a response from the ordering service nodes within 3 seconds, the client would reject the response. In a worse case, for those transactions with ordering latency of 3 seconds, it means that the client rejects those transactions. Our results show an average transaction latency. 
      
\begin{figure}[htbp]
	\centering
	\hspace{-7pt}\includegraphics[width=0.50\textwidth]{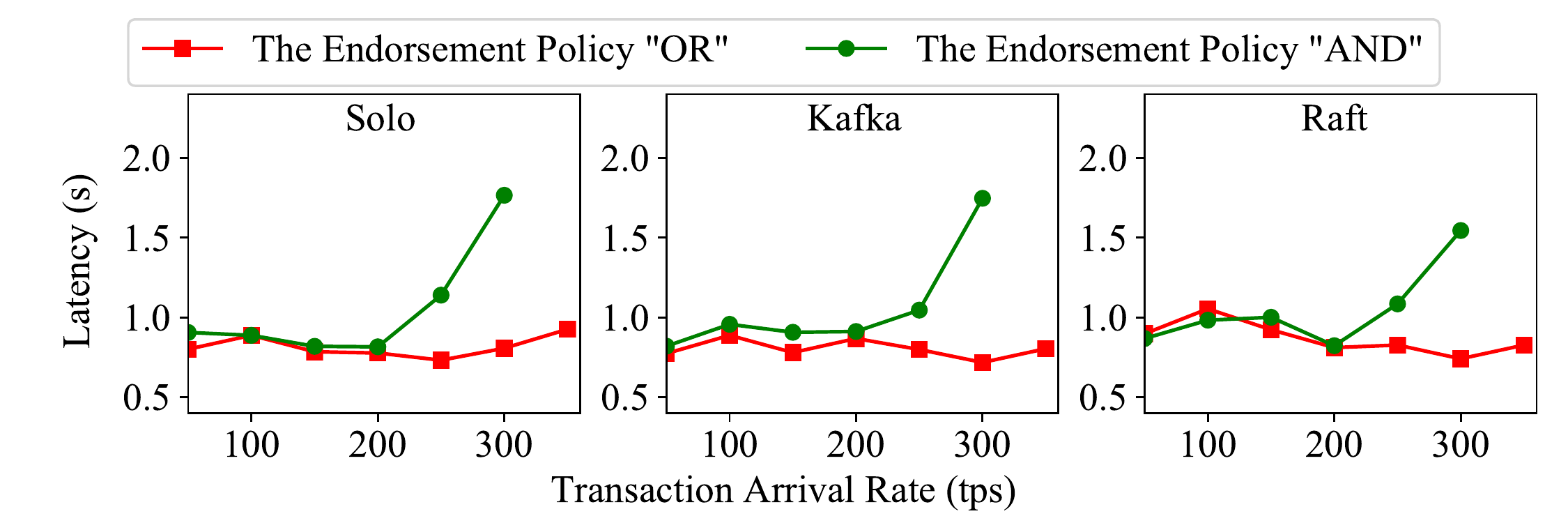}
	\caption{Overall Transaction Latency}
\end{figure}

In summary, an overall transaction latency will increase rapidly when the transaction arrival rate surpasses the peak system performance since many transactions will be rejected. In particular, the peak throughput of the endorsement policy AND comes earlier due to lower peak throughput.

\textbf{Throughput of Each Phase.} Fig. 4 shows a comparison of transaction throughput of each phase, i.e., the execute, order, and validate phase, under the endorsement policy OR. We have two observations. First is that the performance bottleneck occurs in the validate phase. Because the throughput of the execute phase shows a good scalability under the endorsement policy OR, and the workload of the ordering service is not high, that does not need to validate transactions and blocks. Second is that the throughput of each phase grows linearly with the transaction arrival rate before its peak throughput.

\begin{figure}[htbp]
	\centering
	\hspace{-7pt}\includegraphics[width=0.50\textwidth]{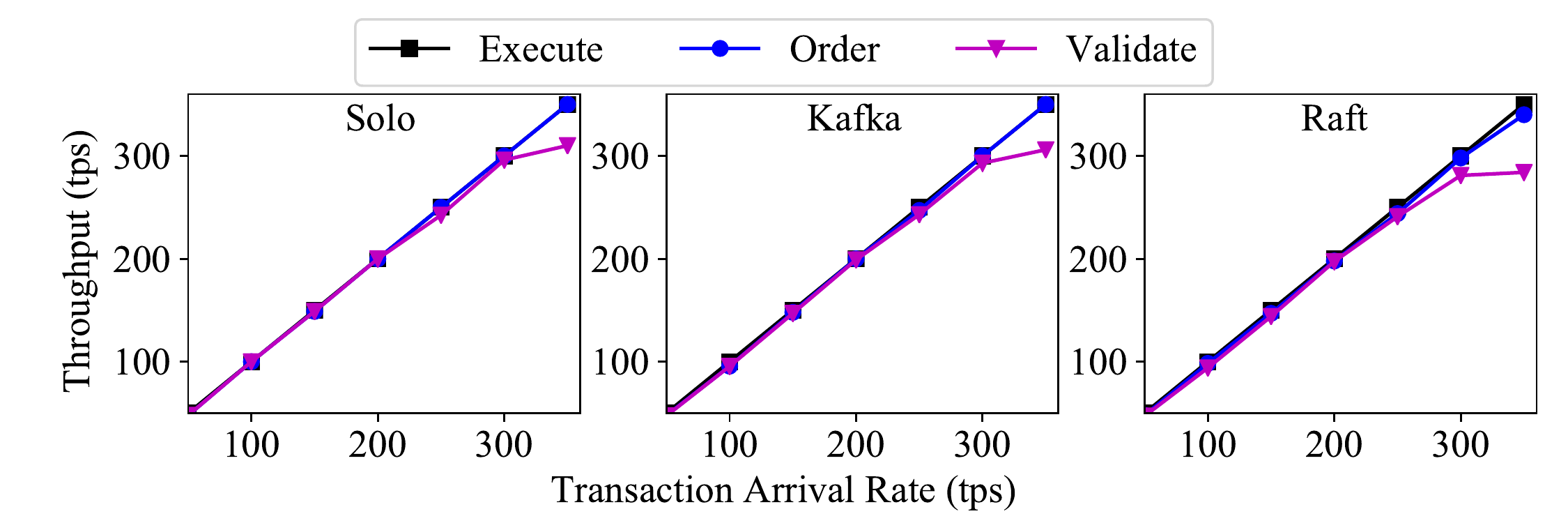}
	\caption{Comparison of Transaction Throughput between the Endorsement Phase and the Ordering Phase under the Endorsement Policy OR}
\end{figure}

\begin{figure}[htbp]
	\centering
	\hspace{-7pt}\includegraphics[width=0.50\textwidth]{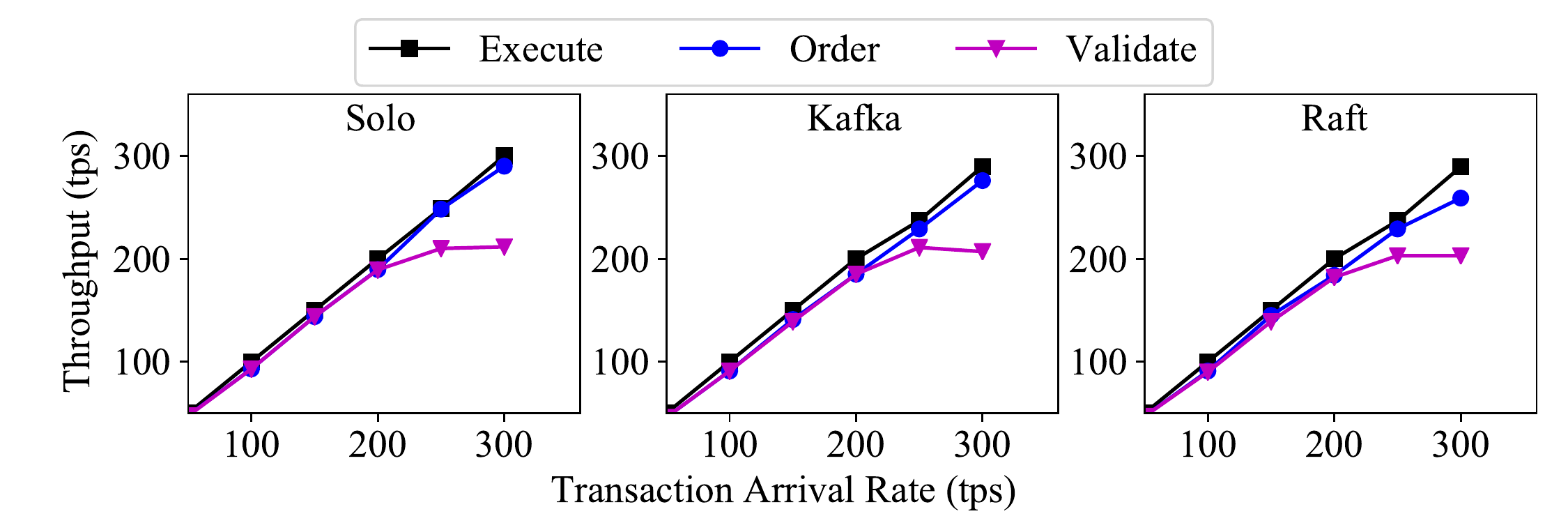}
	\caption{Comparison of Transaction Throughput between the Endorsement Phase and the Ordering Phase with the Endorsement Policy AND}
\end{figure}

Fig. 5 shows a comparison of transaction throughput of each phase under the endorsement policy AND. We have three observations. First is that the throughput scalability under the endorsement policy $\text{AND}_x$ is bad, where $\text{AND}_x$ means to let transaction endorsed by $x$ endorsing peers together. Because the computing resources of the target $x$ endorsing peers are limited. Second is that the performance bottleneck occurs in the validate phase. We can see that the maximum throughput of the validate phase is limited to around 200 tps under the endorsement policy $\text{AND}_5$, where $\text{AND}_5$ means to get transaction endorsed by 5 endorsing peers together. Third is that the throughput of each phase grows linearly with the transaction arrival rate before its peak throughput.

In summary, we find that the throughput scalability under the endorsement policy OR is good, but it is terrible under the endorsement policy AND due to limited computing resources of target peers. Also, the system bottleneck occurs in the validate phase. It is not only because the speed of transaction and block validation is low but also because computing workloads of the validate node are heavy. The node of the execute phase also takes task for validating transactions.

\textbf{Latency of Each Phase.} Fig. 6 shows a comparison of the transaction latency of each phase under the endorsement policy AND. The black line denotes the average latency of the execute phase. The cyan line denotes the average latency of the order and validate phases. We have two observations. First of all, the performance scalability, i.e., latency in this case, under the endorsement policy OR is good. Because we can use more endorsing peers to endorse transactions, a few machines do not limit the computing resources of the execute phase. Secondly, after the workload surpasses the system capacity of the validate phase, Hyperledger Fabric will reject some transactions and result in an increased latency of the order and validate phase.

\begin{figure}[htbp]
	\centering
	\hspace{-7pt}\includegraphics[width=0.50\textwidth]{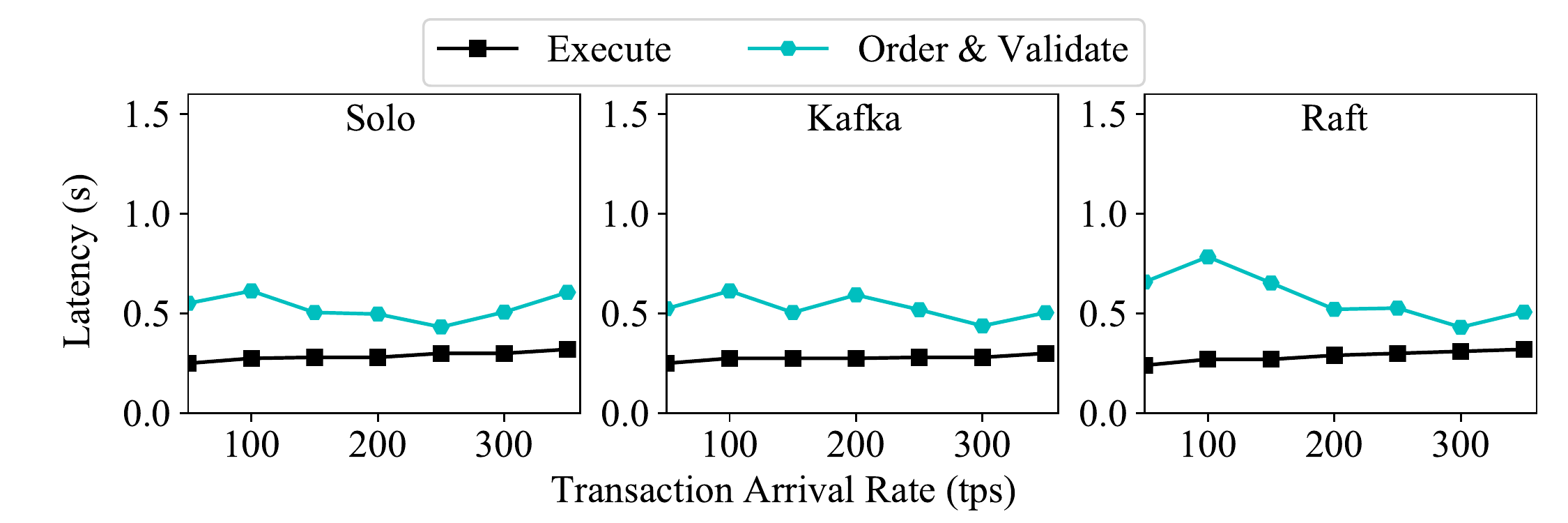}
	\caption{Comparison of Transaction Latency of Each Phase under the Endorsement Policy OR}
\end{figure}

\begin{figure}[htbp]
	\centering
	\hspace{-7pt}\includegraphics[width=0.50\textwidth]{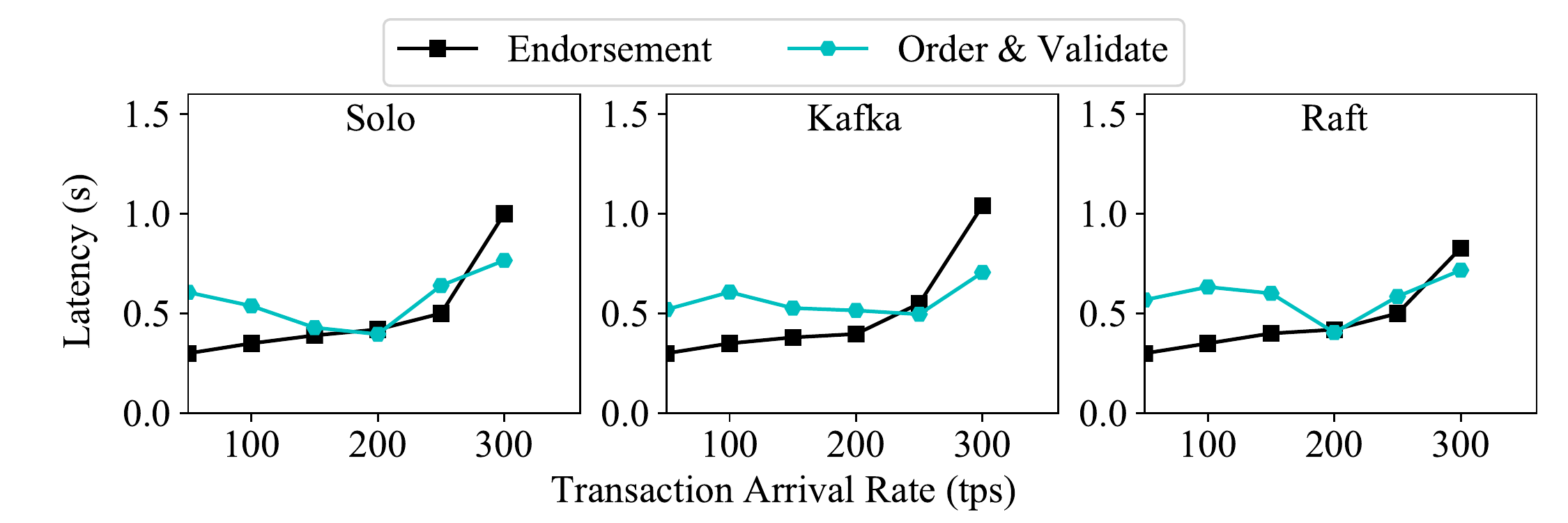}
	\caption{Comparison of Transaction Latency of Each Phase under the Endorsement Policy AND}
\end{figure}

Fig. 7 shows a comparison of latency of each phase under the endorsement policy AND. There are two interesting observations. First of all, the latency of order and validate phases remains stable. Secondly, the latency of all phases sharply grows when the transaction arrival rate surpasses the peak throughput.

\thispagestyle{empty}
In summary, we find that the latency of each phase remains stable before its peak throughput, while it increasingly grows when the transaction arrival rate surpasses its peak throughput due to the queueing effect.

\textbf{Scalability of Endorsing Peers.} Table \uppercase\expandafter{\romannumeral2} shows throughput of different number of endorsing peers. Table \uppercase\expandafter{\romannumeral3} shows the latency of different number of endorsing peers. We observed that the scalability of Hyperledger Fabric much relies on three main factors: endorsement policy and computing capacity of a validation peer. For example, the peak throughput of an endorsement policy OR$_{10}$ can be around 300 transactions per second. The peak throughput of an endorsement policy OR$_3$ can only be 150 transactions per second due to limited target endorsing peers defined by the endorsement policy. The peak throughput of an endorsement policy AND$_5$ can only be around 210 transactions per second due to limited target endorsing peers defined by the endorsement policy and limited computing capacity of validation peers.

\begin{table}[t]
\centering
\caption{Throughput vs. Number of Endorsing Peers}
\scalebox{0.80}{
\begin{tabular}{ccccc}
\hline
\multirow{2}{*}{\begin{tabular}[c]{@{}l@{}}\# Endorsing\\ Peers\end{tabular}} & \multicolumn{4}{c}{Throughput (tps)} \\ \cline{2-5}
                                                                           & OR$_{10}$    & OR$_3$    & AND$_5$    & AND$_3$    \\ \hline
1                                                                          & 50      & 50     & 50      & 50      \\
3                                                                          & 150     & 150    & 150     & 150     \\
5                                                                          & 246     & -      & \textbf{210}     & -       \\
7                                                                          & 310     & -      & -       & -       \\
10                                                                         & \textbf{300}     & -      & -       & -       \\ \hline
\end{tabular}
}
\end{table}

\begin{table}[t]
\centering
\caption{Latency vs. Number of Endorsing Peers}
\scalebox{0.80}{
\begin{tabular}{ccccccccc}
\hline
\multirow{2}{*}{\begin{tabular}[c]{@{}l@{}}\# Endorsing\\ Peers \end{tabular}} & \multicolumn{4}{c}{Execute Latency (s)} & \multicolumn{4}{c}{Order \& Validate Latency (s)} \\ \cline{2-9}
                                                                           & OR$_{10}$     & OR$_3$     & AND$_5$    & AND$_3$     & OR$_{10}$        & OR$_{3}$         & AND$_5$      & AND$_3$      \\ \hline
1                                                                          & 0.25     & 0.25    & 0.3     & 0.285    & 0.551       & 0.551       & 0.55      & 0.55      \\
3                                                                          & 0.28     & 0.28    & 0.39    & 0.38     & 0.505       & 0.505       & 0.43      & 0.43      \\
5                                                                          & 0.3      & -       & \textbf{0.57}    & -        & 0.432       & -           & \textbf{0.7}       & -         \\
7                                                                          & 0.32     & -       & -       & -        & 0.660       & -           & -         & -         \\
10                                                                         & 0.32     & -       & -       & -        & \textbf{0.8}         & -           & -         & -         \\ \hline
\end{tabular}
}
\end{table}

\textbf{Scalability of Ordering Service Nodes.} Fig. 8 shows throughput and latency of Hyperledger Fabric on different number of ordering service nodes. We did not observe different performance of Hyperledger Fabric by adjusting the types of ordering services because the ordering service is not the system bottleneck in our experimental environment. Also, we did not observe a significant difference in latency when scaling the number of ordering service nodes up to 12 either for Kafka or Raft. Similarly, we did not observe a significant difference in latency when scaling the number of ZooKeeper nodes and Kafka Broker nodes up to 7.

\begin{figure}[htbp]
	\centering
	
	\subfigure[\#ZooKeeper=\#Broker=3]{%
		\includegraphics[width=0.23\textwidth]{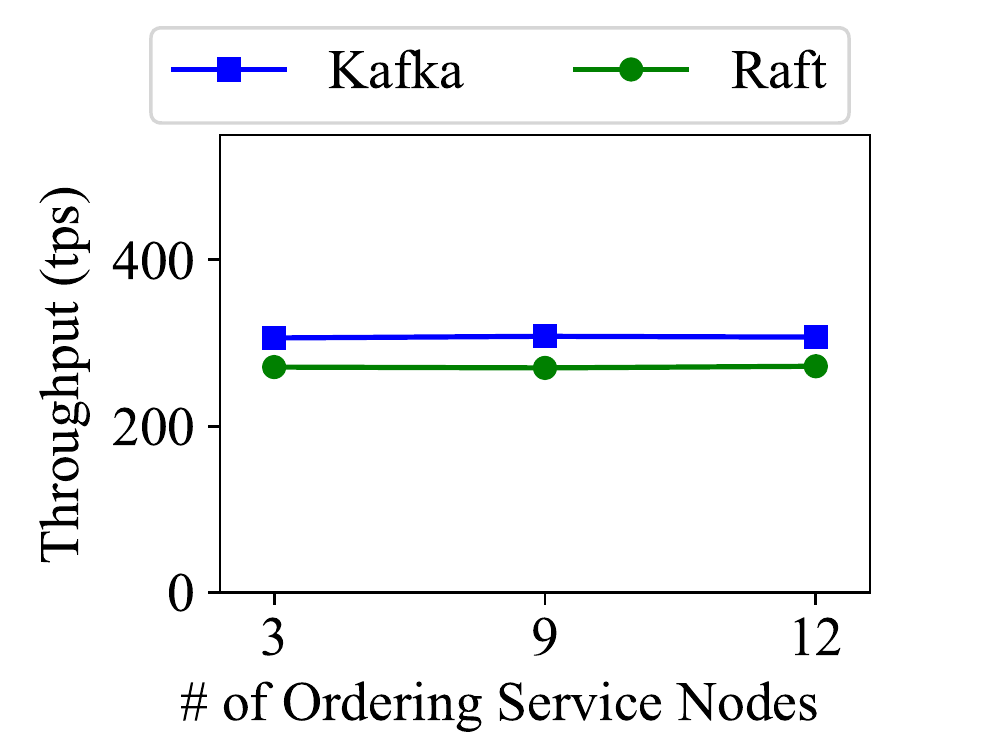}%
	}\hspace{0.2cm}
	\subfigure[\#ZooKeeper=\#Broker=3]{
		\includegraphics[width=0.23\textwidth]{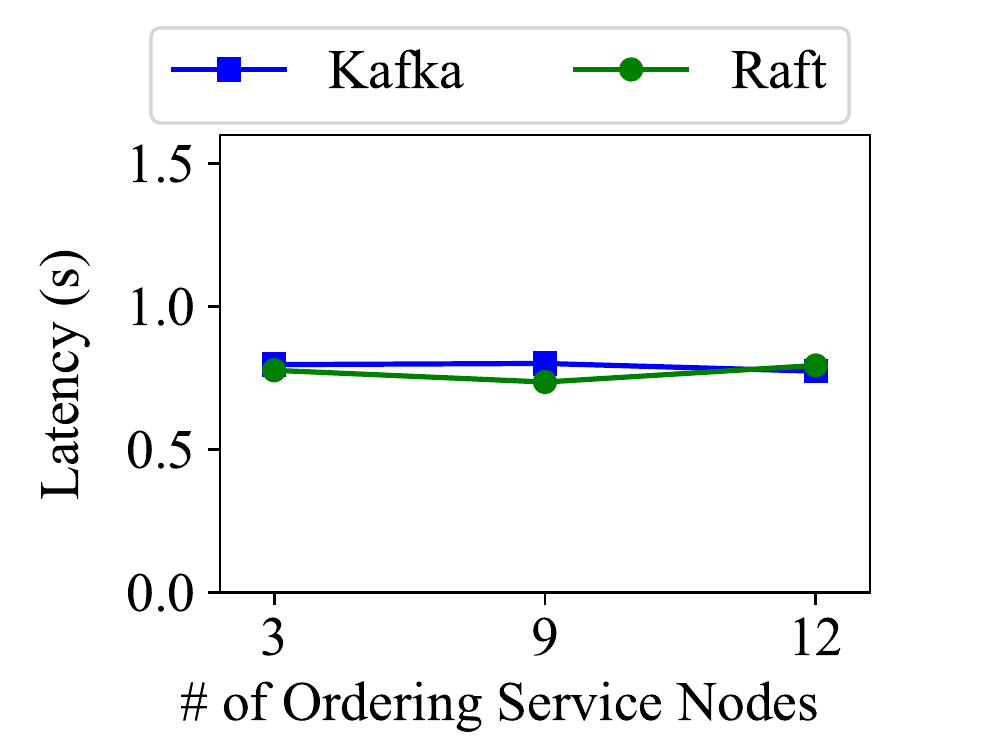}%
	}\hspace{0.2cm}
	\subfigure[\#ZooKeeper=\#Broker=7]{
		\includegraphics[width=0.22\textwidth]{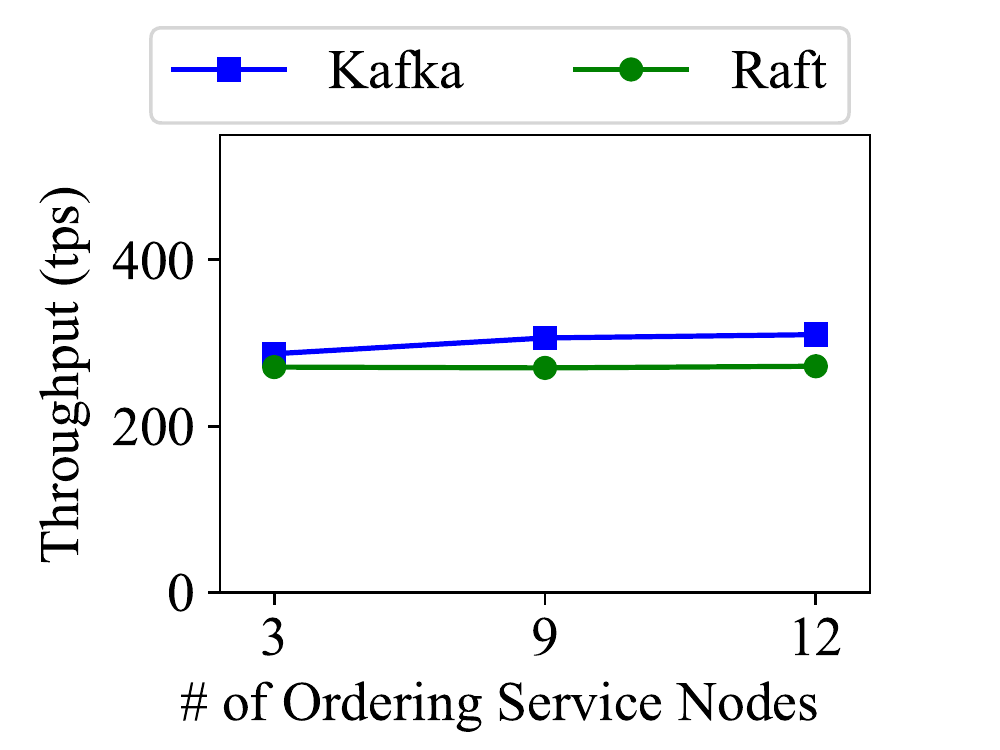}%
	}\hspace{0.2cm}
	\subfigure[\#ZooKeeper=\#Broker=7]{
		\includegraphics[width=0.22\textwidth]{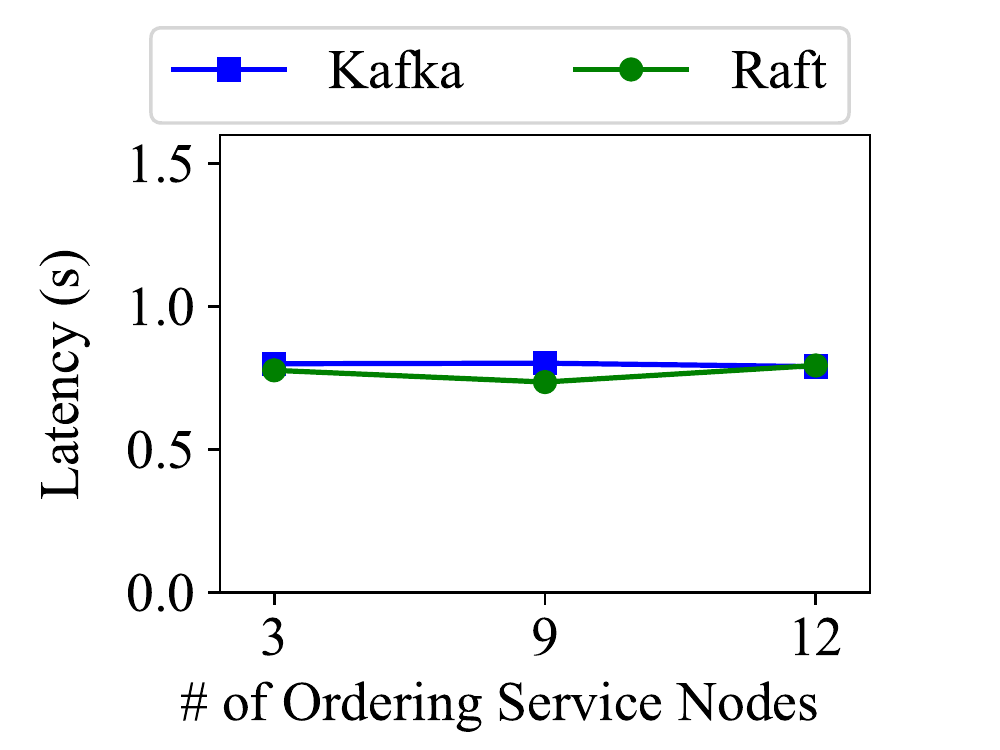}%
	} 
	
	\caption{Throughput (and Latency) vs. Number of Ordering Service Nodes}
\end{figure}

\section{Related Work}

\textit{Workload Designs.} The workload may have different endorsement policy, transaction size, and transaction arrival rate. These factors significantly impact the system performance. For specific application scenarios such as money transfer of bank accounts, the design of the workload also needs to consider the problem of the read-write conflict, though many papers focus on the system-level performance instead of the application-level performance. They \cite{dinh2017blockbench} proposed the first evaluation framework for analyzing private blockchains, with enabling a comparison for different platforms. They \cite{wang2019measurement} studied the blockchain network via measurement and analysis from a specific view. They \cite{baliga2018performance, cilliers2018land} studied the Hyperledger fabric and found that the throughput grows almost linearly with the transaction arrival rate until reaching the system’s peaking processing rate.

\textit{Scalability.} Scalability of Hyperledger Fabric has two types: the scalability of peers and the scalability of the ordering service nodes. The scalability of peers means the effects of peers, including endorsing peers and non-endorsing peers, on system performance. These effects mainly include the communication cost of block propagation between peers. They \cite{androulaki2018hyperledger} showed that the network bandwidth becomes the bottleneck. They \cite{nguyenunderstanding} conducted a performance analysis of Hyperledger Fabric v1.1, in particular, focus on the Kafka ordering service and found that the communication cost of Kafka inner-log replication is small which does not affect the overall throughput. They also found that scaling the Kafka cluster does not affect overall system performance. However, they did not consider the Raft consensus algorithm since it was recently introduced. An overall performance comparison of the ordering services is quite an interesting thing.

\thispagestyle{empty}
\textit{Performance Analysis.} There are many measurement study on Hyperledger Fabric performance, such as \cite{androulaki2018hyperledger,baliga2018performance}. A noting work that aims to model the Hyperledger Fabric Performance is \cite{sukhwani2017performance}. They proposed a Stochastic Reward Nets (SRN) to study the consensus time for the Hyperledger Fabric system. However, there are some limitations. First of all, they used an older version of the Hyperledger Fabric v0.6, which has a considerable difference with a newer version of Hyperledger Fabric v1.x. There is no concept of the ordering service in v0.6. Second, their work only analyzes the transaction latency from the view of the client while showing limitations in the more detailed latency.

\section{Conclusion}

We presented a performance study and bottleneck analysis on Hyperledger Fabric blockchain platform. We studied the performance characterization of each phase of the transaction life cycle. We also made a comparison of different ordering services. The experiment results showed some interesting findings. For example, the execute phase showed a good performance scalability under the OR endorsement  policy, but it was exacerbated under the AND endorsement policy. The validate phase was the system bottleneck on our testbed. It is not only because the speed of transaction and block validation is low, but also because the computing workload of the validate node is heavy. Overall, our work helps to better understand the performance characteristics and bottleneck of Hyperledger Fabric.

\bibliographystyle{IEEEtran}
\bibliography{ref}

\end{document}